\documentclass[a4paper]{jpconf}
\usepackage{graphicx}
\usepackage{amsmath}

\begin{document}
\title{Distance Estimate of Tycho's SNR}

\author{A V Kozlova $^{1,2}$, S I Blinnikov $^{3,4}$ }

\address{$^{1}$ Ioffe Institute, 194021 St. Petersburg, Russia}
\address{$^{2}$ Sternberg Astronomical Institute, Moscow M. V. Lomonosov State University, 119234 Moscow, Russia}
\address{$^{3}$ Institute for Theoretical  and Experimental Physics (ITEP), 117218 Moscow, Russia}
\address{$^{4}$ All-Russia Research Institute of Automatics (VNIIA), 127005 Moscow, Russia}

\ead{ann\_kozlova@mail.ioffe.ru}

\begin{abstract}
We present an approach to finding distances to young supernova remnants. Our method is based on hydrodynamical simulations of Tycho's SNR using the SUPREMNA code. For the explosion models, we use the classical W7 deflagration model and the delayed detonation model that was previously shown to provide good fits to the X-ray emission of Tycho's SNR. Combining our hydrodynamical simulations with the Chandra high-resolution images of the remnant obtained in 2015, we derive a distance estimate to Tycho's SNR of $2.8\pm 0.4$ kpc. 
\end{abstract}

\section{Introduction}

Supernova remnants (SNRs) play a significant role in the enrichment of the interstellar medium (ISM) with heavy elements, they are a source of energy injection and likely the sources of Galactic cosmic rays. Studying these objects is essential for our understanding a wide range of sections in physics and astronomy, for example, such important field as supernova explosion mechanisms, cosmological parameters, and distances to host galaxies (in case of Type Ia SN). Reliable distance determination to SNRs is a key in constraining their physical parameters like the size, the age and the explosion energy, which helps us to study their environment and ISM.

Tycho's supernova remnant is a member of a small subclass of Galactic SNRs 
known as the ``historical supernovae". It is the remnant of a standard 
Type Ia \cite{Krause} (i.e., a thermonuclear explosion of a white dwarf star in a close binary) SN  
observed by Tycho Brahe in 1572. The remnant has been widely studied across the 
electromagnetic spectrum (radio: \cite{Dickel}, optical: \cite{Ghavamian}, X-rays: \cite{Sato}, $\gamma$-rays: \cite{Archambault}). However, the distance remains uncertain with a spread of published values mostly between 2 and 4 kpc \cite{Hayato}.  

The rest of this paper is organized as follows. In Section 2, we report on details of modeling the X-ray emission from Tycho's SNR. We present the results and compare them with {\it{Chandra}} observations in Section 3. Finally, in Section 4, we conclude with a few remarks.

\section{Simulations}
Our goal is to model the X-ray emission from the Tycho's SNR using the grid of synthetic spectra computed with {{SUPREMNA}} \cite {Sorokina}. It is a one dimensional spherically symmetric hydrodynamical code, which is aimed at the macroscopic scale study of SNRs. The code uses an implicit Lagrangean formulation and takes into account the most important physical processes, such as time-dependent ionization, the influence of radiative losses, the account of electron thermal conduction and nonthermal particles \cite{Kosenko}. We attempt to reproduce the observed X-ray emission, calculate the SNR dynamics, and derive distance estimate by comparing shock velocity with the observed expansion rate. 

We base our ejecta models on two Type Ia SN explosion models: a classical deflagration W7 model \cite{Nomoto}, which results from a thermonuclear, subsonic flame propagation through the white dwarf, and a modern delayed detonation  ``deldet" model \cite{Ropke, Seitenzahl}, which assumes that the flame starts as a deflagration and turns into a detonation later on, as they appear to give the good overall agreement with the general characteristics of SN and well reproduce the X-ray emission of Tycho's SNR. For W7 explosion we use 300 zones: 99 for the ejecta and the rest for the ISM; for ``deldet" model the hydrodynamical mesh consists of 200 zones, 100 of which is for the ejecta. We expand the original explosion models up to the age of 15--35 years and then start the simulations of an SN expansion into uniform ISM. During the calculation, we record the history of velocity, the temperature, the density and ion composition changes for each mesh zone. The resulting ionization stages for 15 most abundant elements (for which the calculations of kinetics were performed in the code) at the age of Tycho's remnant are used to evaluate an X-ray spectrum. 

The synthetic spectra are mainly characterized by the SN explosion model, the age of the SNR, the density of the ISM, and non-Coulomb energy exchange between electrons and ions ($q$). Of the four parameters, the age is known to be 443 yr for the Tycho's SNR (at the time of {\textit{Chandra}} observation used). Thus, only two remaining parameters can be varied for each of explosion models. We performed our hydrodynamical calculations for different values of the ISM density in a $(0.5 - 5)\times 10^{-24}$~g/cm$^{3}$ range. The parameter $q$ specifies the fraction of artificial viscosity $Q$, added to the pressure of ions ($(1-q)Q$ added to electrons). If only the collisional energy exchange is taken into account, then $q=1$ and we get the usual system of equations with the heating of only ions at the front. The detailed description of this approach is presented in \cite{Sorokina, Kosenko}. We have sampled this parameter space with three points (0.5, 0.8, 1) to see if there is an appreciable non-Coulomb energy exchange due to the presence of magnetic field and plasma instabilities. Calculated spectra were converted to XSPEC table models with this interpolation parameters and fitted to {\textit{Chandra}} X-ray observation. Then with the best-fitting values for interpolation parameters a new hydrodynamical model was calculated.

\section{Results and Comparison with observation}
\begin{figure}[h]
\center\includegraphics[width=0.8\textwidth]{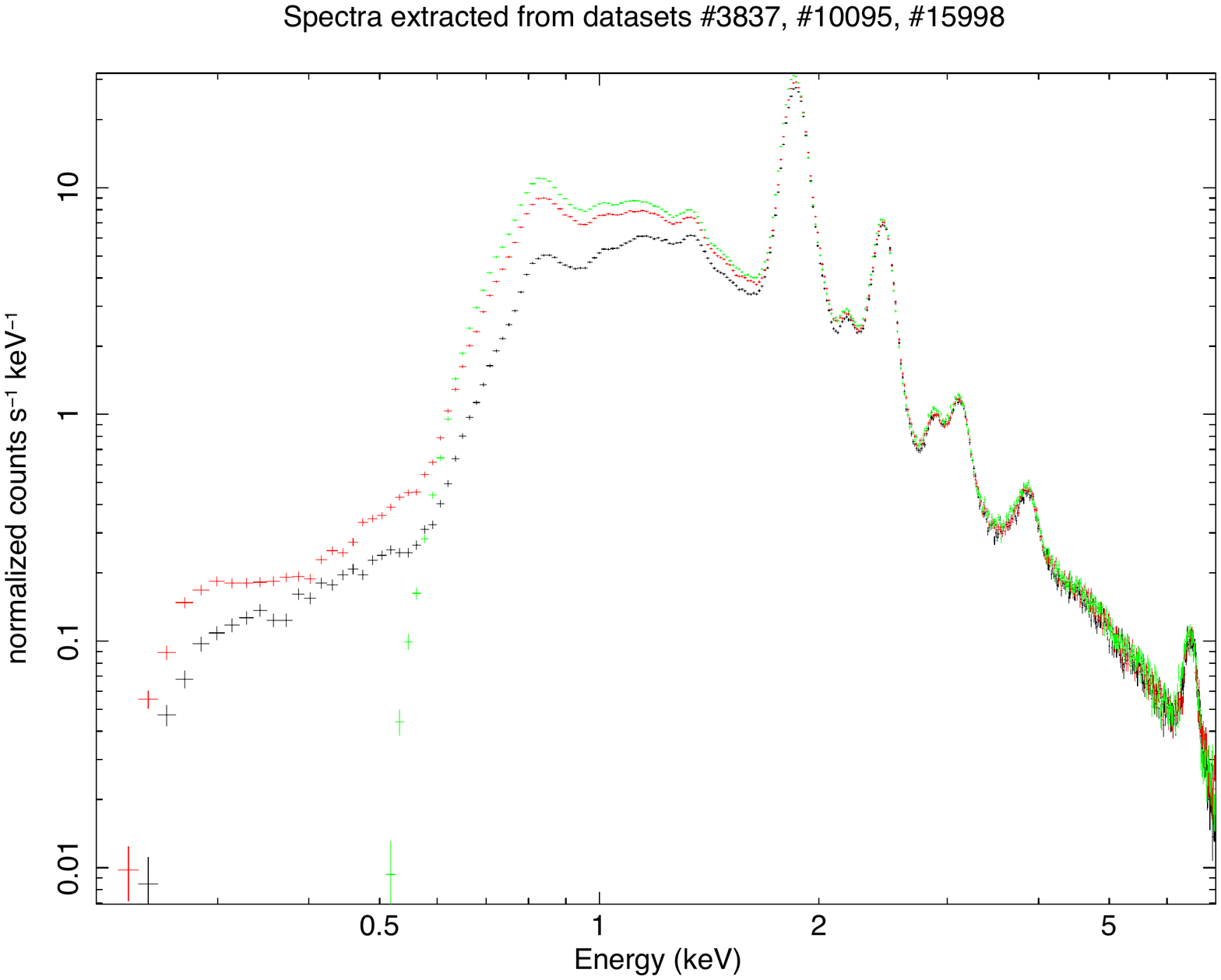}
\caption{\label{comparingspectra} The background-substracted spectra of the western part of Tycho, extracted from Obs. ID 3837 (green), Obs. ID 10095 (red), and Obs. ID 15998 (black). }
\end{figure}

{\it{Chandra}} archival data over a 15-year baseline from 2000 to 2015 are available for Tycho's SNR. We obtained data from three observations, with the longest exposure times: Obs. ID 3837 (146 ks, taken on 29 April 2003), Obs. ID 10095 (173 ks, taken on 23 April 2009), and Obs. ID 15998 (147 ks, taken on 22 April 2015). To analyze datasets we used the CIAO 4.8 and CALDB 4.7.2 \cite{Fruscione}. We reprocessed {\it{Chandra}} data with {\texttt{chandra\_repro}} and extracted source and background spectra using {\texttt{specextract}}. Since our code is one dimensional, we focus on the western part of Tycho, avoiding irregular ejecta clumps in the southeast and the interactions with a denser ISM toward the northeast \cite{Decourchelle, Reynoso}. The background-substracted spectra of the same region, which we choose to study, from three different datasets are shown in Figure \ref{comparingspectra}. Differences in the energy spectra are almost absent at $E\geq 1.5$~keV, so below we only consider the analysis of latest available observation (Obs. ID 15998). 

\begin{figure}[h]
\begin{minipage}{20pc}
\center\includegraphics[width=16pc]{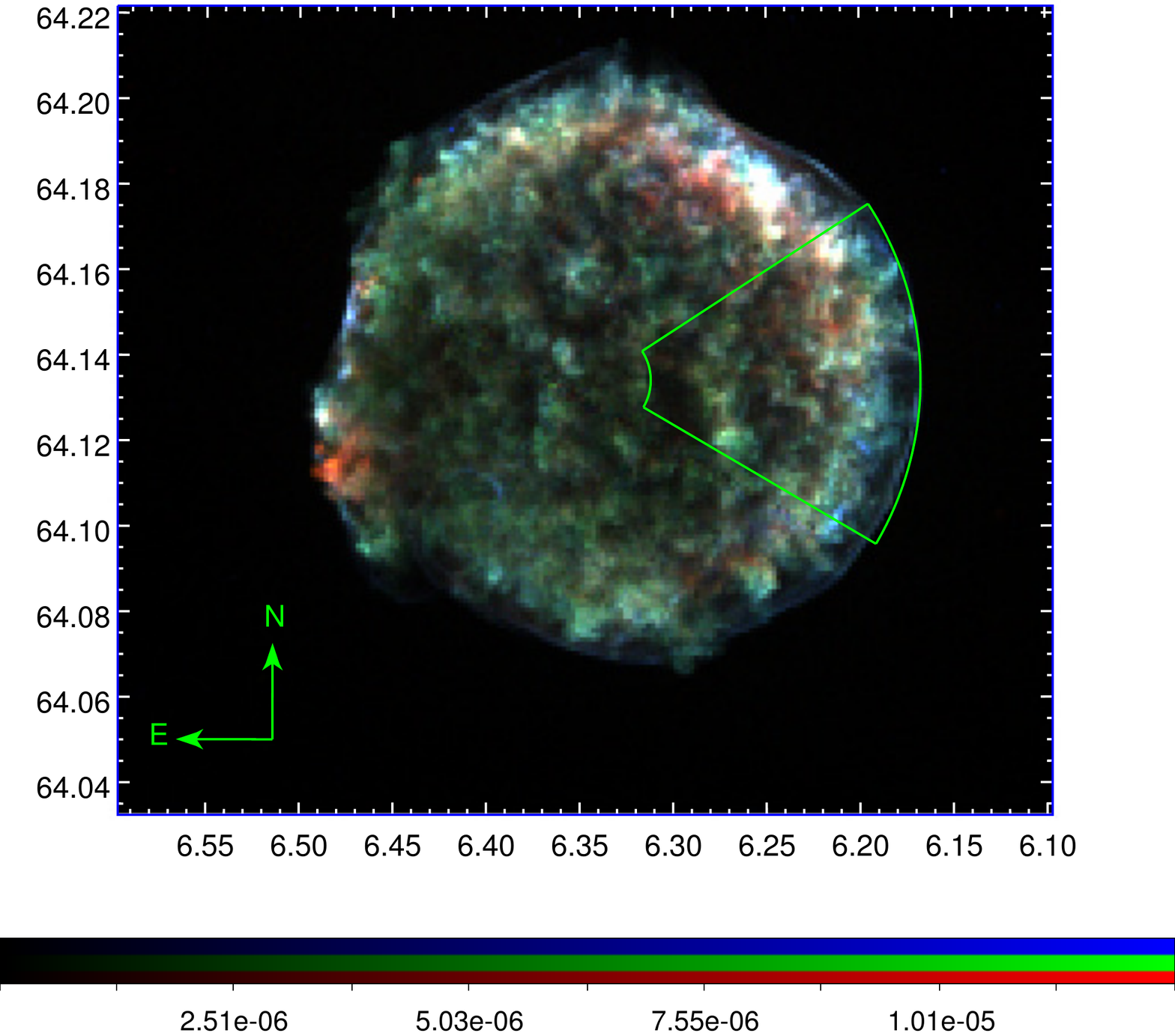}
\end{minipage}
\hfill
\begin{minipage}{19pc}
\includegraphics[width=19pc]{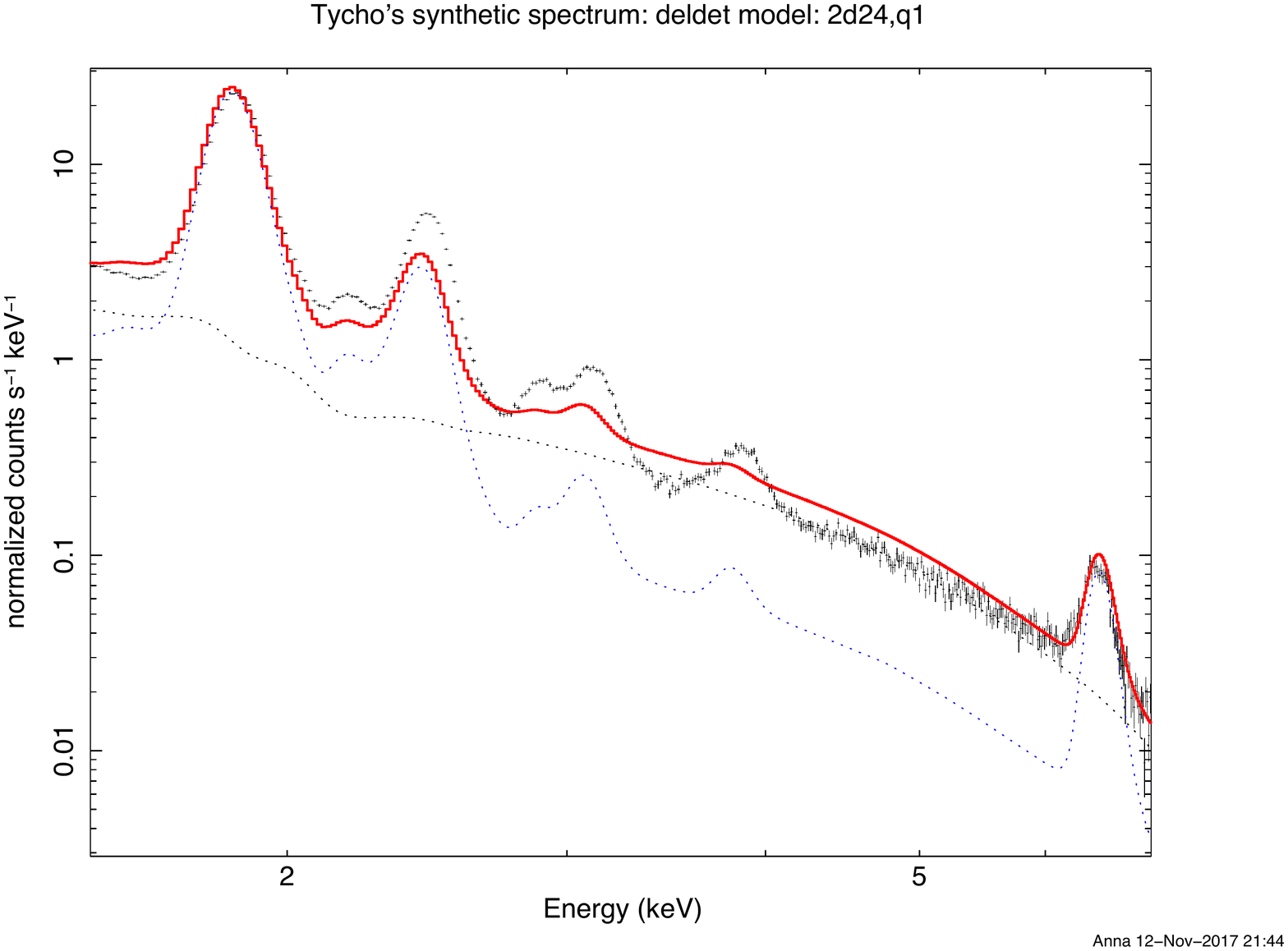}
\end{minipage}
\vfill
\begin{minipage}{19pc}
\includegraphics[width=19pc]{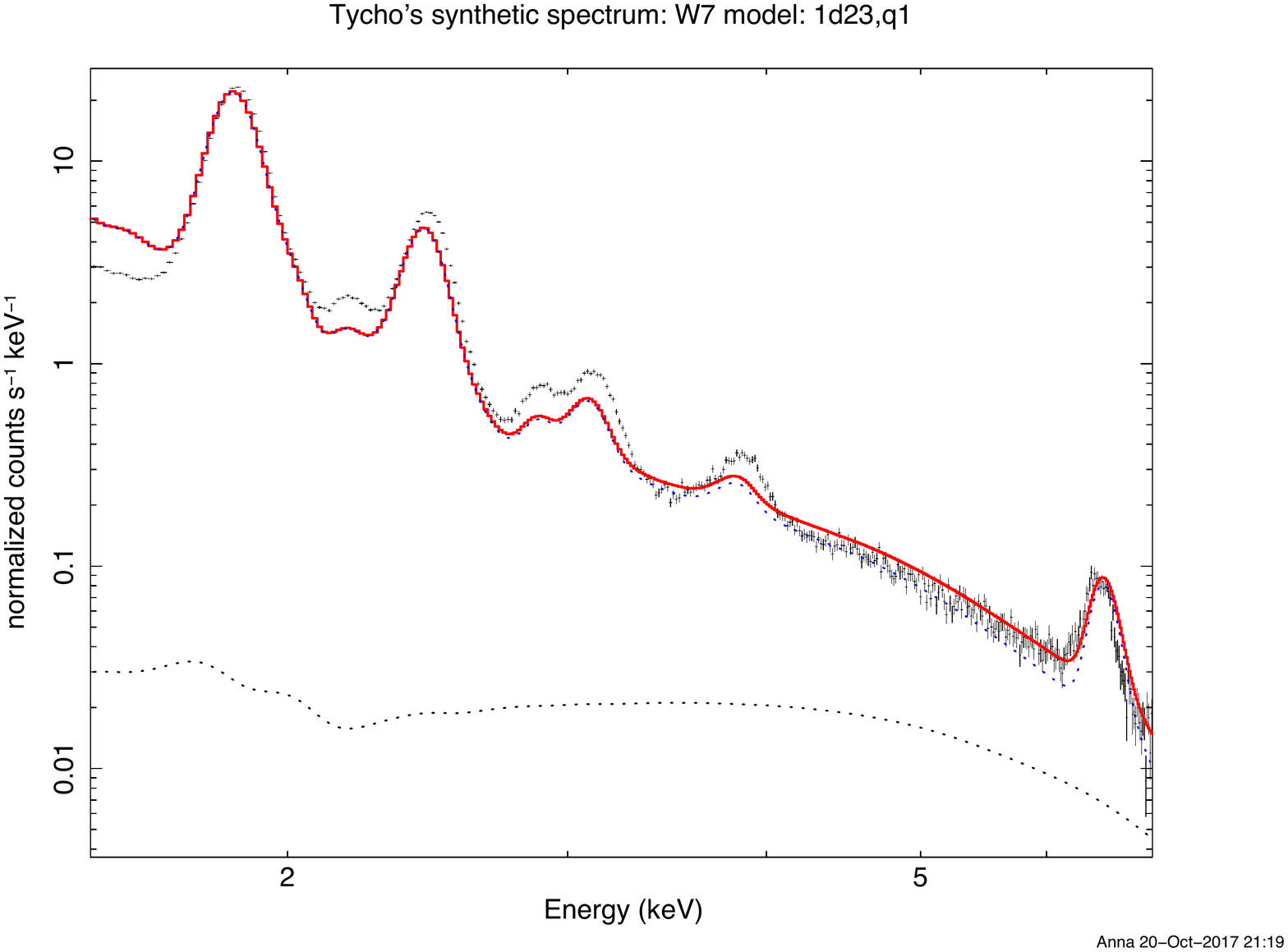}
\end{minipage}
\hfill
\begin{minipage}{19pc}
\centering
\includegraphics[width=19pc]{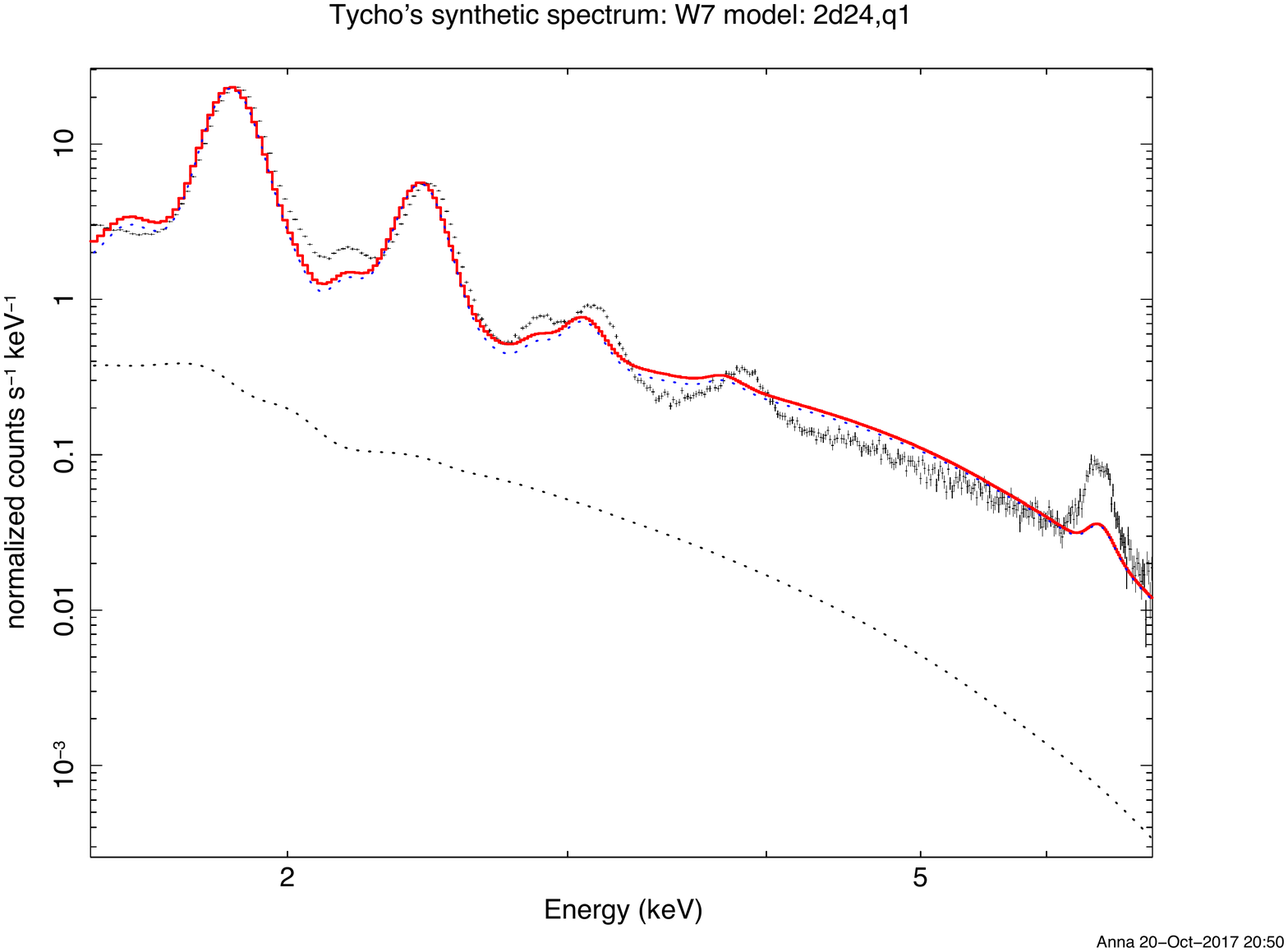}
\end{minipage} 
\caption{\label{regionandspectra}{\it{Top row, left: }} A three-color {\it{Chandra}} image of the Tycho's SNR obtained from the observation ID 10095, with the source spectrum extraction region indicated in green. Red corresponds to low-energy ($0.5 - 1.2$~keV), green to middle-energy ($1.2 - 2.0$~keV), and blue to high-energy ($2.0 - 7.0$~keV). Comparison between the ejecta models and the X-ray spectrum of region. ``deldet" model, ISM density $\rho=2 \times 10^{-24}$~g/cm$^{3}$ ({\it{Top row, right}}). W7 model, $\rho=2 \times 10^{-23}$~g/cm$^{3}$ ({\it{bottom row, left}}) and W7 model $\rho=2 \times 10^{-24}$~g/cm$^{3}$ ({\it{bottom row, right}}). The power-law component is displayed (dotted line).}
\end{figure}

We used XSPEC 12.9.0 \cite{Arnaud} for spectral analysis, and performed model fitting in the 1.5-7.0 keV band. We choose this band in order to avoid the uncertain  absorption features from interstellar matter, and, at the same time, to include in our analysis a Si line at $\sim 1.86$ keV. Results of the spectra simulations for different explosion models are shown in Figure \ref{regionandspectra}.

Taking into account the simplicity of the models, limitations of one-dimensional calculations, and the uncertainties in the atomic data, we do not expect a valid fit from the statistical point of view, but nevertheless, we report the value of reduced $\chi^2$ as an indicator of goodness of fit. We find that the W7 model does not have enough Fe to reproduce the Fe K$\alpha$ flux in the Tycho. This model gives a good fit to X-ray spectra of Tycho ($\chi^2 =56.6$), only when it requires the high ISM density of $\sim 10^{-23}$~g/cm$^3$, which is not in agreement with the observed properties of the remnant \cite{Katsuda}. The model based on ``deldet" explosion matches well with Tycho X-ray spectra ($\chi^2 =63.7$) and reproduces the observed parameters. Thus, we choose the ``deldet" model to be the preferred one. Note that we add a power-law component to account for X-ray synchrotron radiation contribution to the spectrum because there is no self-consistent theory for it to implement in the code \cite{Cassam}.

Using the predicted by SUPREMNA shock velocity $v_s$ and the proper motions $\mu$ on the western side of the remnant derived by Williams \cite{Williams16}, the distance to the remnant was evaluated: 
$$
d (\text{parsecs}) = \frac{v_s (\text{km/s})}{4.74 \mu (\text{arcsec/yr})}
$$
\begin{center}
\begin{table}[h]
\centering
\caption{\label{distance}Estimates of the distance to Tycho's SNR.}
\begin{tabular}{cc}
\br
Model name & Distance (kpc)\\
\mr
``deldet" & $2.8\pm 0.4$ \\
W7, $\rho_{\text{ISM}} \sim 10^{-24} $g/cm$^3$ & $2.6\pm 0.4$\\
W7, $\rho_{\text{ISM}} \sim 10^{-23} $g/cm$^3$ & $1.8\pm0.3$ \\
\br
\end{tabular}
\end {table}
\end{center}

The calculated distances are summarized in Table \ref{distance}.

\section{Conclusion}
We have modeled the evolution, performed a hydrodynamical calculation of X-ray spectrum of Tycho's SNR and compared the model X-ray emission with archived {\it{Chandra}} observation. We have found the delayed detonation model, expanding into a uniform ISM of $\rho  = 2 \times 10^{-24}$~g/cm$^{3} $ to be the preferred one for Tycho's SNR. This is in agreement with results of \cite{Badenes}. Predicted by SUPREMNA shock velocities are slightly higher than those obtained previously by Fang et al \cite{Fang} in a 3D hydrodynamical simulation. This difference is mainly due to the density configuration used in simulations (uniform medium of constant density vs a latitude-dependent wind-blown cavity).  Based on ``deldet"  model the distance to the SNR is estimated to be $2.8\pm 0.4$ kpc. This estimate is in agreement with other shock-wave methods \cite{Hayato}. 

The errors can be improved with the application of more detailed models to different regions of the remnant. Our method of distance determination is more robust than other shock-wave methods which rely on line emission of knots which have irregular structure. 

In the future, we plan to modify SUPREMNA code for self-consistent treatment of cosmic rays.

\ack
The work by A. K. (modeling X-ray emission of Tycho's SNR and deriving distance estimates) is supported by the RSF grant 16-12-10519. We thank Ping Zhou for useful references and discussion. 

\section*{References}

\bibliographystyle{iopart-num}
\bibliography{example}

\end{document}